%% file: arxiv.tex
\documentclass[a4paper,12pt]{article}
\title{Biomolecular System Energetics}

\usepackage[noblocks]{authblk}
\author[1]{Peter J. Gawthrop\footnote{Corresponding
    author. \textbf{peter.gawthrop@unimelb.edu.au}}}
\author[1,2,3]{Edmund J. Crampin}
\affil[1]{
  Systems Biology Laboratory,
  Department of Biomedical Engineering,
  Melbourne School of Engineering,
  University of Melbourne,
  Victoria 3010, Australia.
   }
\affil[2]{School of Mathematics and Statistics, University of
      Melbourne University of Melbourne, Victoria 3010,
  Australia}
\affil[3]{School of Medicine, University of Melbourne, Victoria 3010,
  Australia}

\newcommand{\HalfWidth}{0.45}
\newcommand{\FullWidth}{0.6}

\usepackage{a4wide,times,siunitx}
 
\include{BG-sym}
\include{BG-maths}
\include{BG-chem}

\usepackage[numbers,square,sort,longnamesfirst]{natbib}
\usepackage[hidelinks]{hyperref}
\usepackage{url,doi}

\usepackage{graphicx,subfigure,fancybox,color}
\newcommand{\Fig}[2]{
  \includegraphics[width=#2\linewidth]{#1.pdf}
   \label{subfig:#1}
}

\newcommand{\SubFig}[3]{
  \subfigure[#2]{
    \includegraphics[width=#3\linewidth]{#1.pdf}
    \label{subfig:#1}
  }
}

\usepackage{epigraph}
\newcommand{\Ch}[1]{
\ch[label-style=\tiny]{#1}
}

\begin{document}

\maketitle
\begin{abstract}
  Efficient energy transduction is one driver of evolution; and thus
understanding biomolecular energy transduction is crucial to
understanding living organisms. As an energy-orientated modelling
methodology, bond graphs provide a useful approach to describing and
modelling the \emph{efficiency} of living systems. This paper gives
some new results on the efficiency of metabolism based on bond graph
models of the key metabolic processes: glycolysis.
\end{abstract}
\tableofcontents
\newpage
\section{INTRODUCTION}
\label{sec:introduction}
\epigraph{ \emph{Katchalsky's \index{Katchalsky} breakthroughs in
    extending bond graphs to biochemistry are very much on my own
    mind. I remain convinced that BG models will play an increasingly
    important role in the upcoming century, applied to chemistry,
    electrochemistry and biochemistry, fields whose practical
    consequences will have a significance comparable to that of
    electronics in this century.}  }{Henry Paynter, 1993} 

As noted by \citet{Pay93a}, \citet{OstPerKat71} used bond graphs in
their seminal paper \emph{Network Thermodynamics} to describe and
analyse systems of coupled chemical reactions. This work was extended
by \citet{Kar90}, \citet{Cel91}, \citet{ThoMoc06} and
\citet{GreCel12}.
These ideas were introduced to the Systems Biology community by
\citet{GawCra14,GawCra16}. As noted by \citet{Kar90} the bond graph
approach is particularly appropriate to electrochemical systems and
therefore can be used to model the bioenergetics of excitable
membranes \citep{GawSieKam17,PanGawTra18X}, redox reactions and
chemiosmotic energy transduction in mitochondria \citep{Gaw17a}.

Organisms need energy to drive essential organs including the brain
\citep{SteLau15,Niv16}, heart \citep{Neu07,Kat11} and muscles
\citep{SmiBarLoi05}.  As discussed in the text books
\citep{BerTymGat15,AlbJohLew15}, this energy is derived from
metabolism involving glycolysis%
\footnote{Glycolysis is the metabolic process converting the sugar
  Glucose to the intermediate high-energy molecule pyruvate.}%
, the TCA cycle%
\footnote{The tricarboxylic acid (TCA) cycle, also known as the citric
  acid cycle or the Krebs cycle, converts the high-energy molecule
  pyruvate into the high-energy molecule NADH (reduced nicotinamide
  adenine dinucleotide).}  and the mitochondrial%
\footnote{Mitochondria are organelles within many cells which provide
  efficient conversion of the products of the TCA cycle into ATP.}
respiratory chain \citep{NicFer13}. Both glycolysis and the
mitochondrial respiratory chain produce energy storage molecule ATP
(adenosine triphosphate)
Energy plays a key role in evolution \citep{NivLau08,Lan14}. In
particular, evolutionary pressure would be expected to lead to
organisms with both efficient energy production and consumption.

Efficiency of production has been experimentally investigated in the
context of glycolysis and the TCA cycle by \citet{ParRubXu16} and in
the context of the mitochondrial respiratory chain by
\citet{LarTorLin16}.
Efficiency of energy consumption has been considered in the
context of neurons by \citet{Niv16}, in the
context of the heart by \citet{LopDha14}, and in the context of muscle
by \citep{SmiBarLoi05}.
Feedback systems regulate metabolism and its efficiency
\citet{Har08,TraLoiCra15,DonSanLin17}.

In the context of living systems, efficiency has been defined in a
number of ways including ATP/O ratio \citep{LarTorLin16} and
thermodynamic definitions consistent with engineering practice
\citep{SmiBarLoi05,Nat16}. The latter approach is used here.

As discussed by \citet{Bea11} meaningful numerical simulation of
living systems requires, \emph{inter alia}, a firm thermodynamic
foundation. Such a foundation is especially important in the context of
investigating efficiency. As an energy-based modelling approach, the
bond graph provides a firm foundation for studying living systems in
general and biomolecular system energetics in particular.

\S~\ref{sec:modelling} is an introduction to bond graph modelling of
biomolecular systems based on the specific system analysed in the
paper.
\S~\ref{sec:efficiency} gives a bond graph  approach to the efficiency
of biomolecular systems illustrated by the example of glycolysis and
\S~\ref{sec:conclusion} concludes the paper.


\section{MODELLING}
\label{sec:modelling}
This section introduces the modelling of biomolecular systems using
bond graphs using the example of the first stage of human metabolism,
glycolysis, which converts the high-energy molecule glucose (\ch{GLC}) to the
high-energy molecule pyruvate (\ch{PYR}), and also generating
adenosine triphosphate (\ch{ATP}) and (reduced) nicotinamide adenine
dinucleotide (\ch{NADH}). 
\S~\ref{sec:chemical-equations} looks
at a single reaction: \ch{ATP} hydrolysis,
\S~\ref{sec:numerical-values} discusses how the numerical values were
obtained and
\S~\ref{sec:redox-reactions} examines \emph{redox}
(reduction-oxidation) reactions.
\S~\ref{sec:glycolysis} discusses the modular bond graph modelling of
glycolysis: the first stage of aerobic respiration.

\subsection{Chemical reactions}\label{sec:chemical-equations}
\begin{figure}[htbp]
  \centering
  \Fig{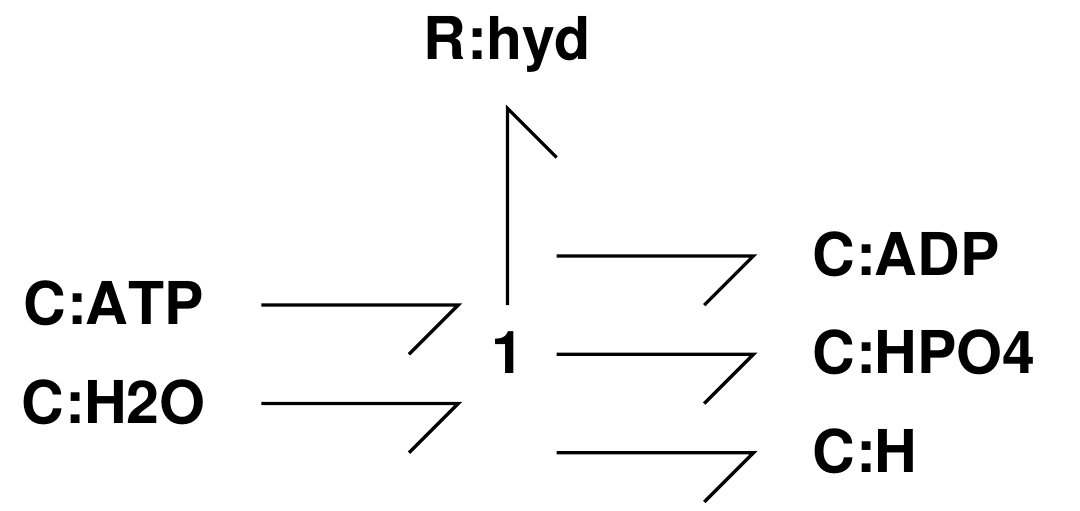}{\HalfWidth}
  \caption{\ch{ATP} hydrolysis: reaction (\ref{eq:ATP})}
  \label{fig:ATP}
\end{figure}
The reaction of adenosine triphosphate \ch{ATP} with water \ch{H2O} to
form adenosine diphosphate \ch{ADP}, inorganic phosphate \ch{HPO4^2-} and
a proton \ch{H+} is known as \ch{ATP} \emph{hydrolysis} and is given by
\citet[\S~18.4, p.564]{BerTymStr12} as
\begin{equation}\label{eq:ATP}
\ch{ATP + H2O <>[hyd] ADP  + HPO4 + H}
\end{equation}
In bond graph terms, each chemical (\ch{ATP}, \ch{H2O} etc.) can be
regarded as a \C component accumulating each particular chemical and
the hydrolysis reaction \texttt{hyd} can be regarded as an \R
component driven by the chemical potentials
$\mu_{\ch{ATP}}, \mu_{\ch{H2O}}$ etc. with units of \si{J/mol} generating
the molar flow $v$ with units of \si{mol/sec}. Reaction stoichiometry
implies that the molar flow $v$ is out of \ch{ATP} and \ch{H2O} and
into \ch{ADP}, \ch{HPO4} and \ch{H}. As the the product $\mu
\times v$ has units of \si{J/sec}, $\mu$ and $v$ are covariables.
Hence the reaction (\ref{eq:ATP}) may be modelled
by the bond graph  of Figure \ref{fig:ATP}.

As discussed by \citet{Gaw17a}, it is helpful (to engineers) to measure quantity in
Coulombs rather than moles and the corresponding conversion factor is
Faraday's constant $F=\SI{96485}{C.mol^{-1}}$. Noting that
\si{J.C^{-1}} has the special unit Volt (\si{V}) and \si{C.s^{-1}} has
the special unit Ampere (\si{A}), the effort covariable becomes the
\emph{Faraday-equivalent potential} $\phi = F\mu\si{V}$ and the flow
covariable becomes the \emph{Faraday-equivalent flow}
$f = \frac{1}{F}v\si{A}$.

Using the standard formula for chemical potential as a function of
quantity \citep{AtkPau11},
the \emph{Constitutive Relationship} (CR) of a chemical \C component associated with substance A gives
the potential $\phi_A$ in terms of the amount $x_A$ 
in terms of the potential $\phi^\Std_A$ and amount $x_A^\Std$ at standard conditions 
\begin{align}
  \phi_A &= \phi_A^\Std + V_N \ln \frac{x_A}{x_A^\Std}\label{eq:phi_A}\\
         &= V_N \ln K_A x_a \label{eq:phi_A_1}\\
  \where V_N &= \frac{RT}{F} \approx \SI{26}{mV}
  \text{ and } K_A = \frac{\exp \frac{\phi_A^\Std}{V_N}}{x_A^\Std}
\end{align}
The Faraday-equivalent potential $\phi^\std_A$ at any other operating
point can be computed from
Equation \eqref{eq:phi_A} as
\begin{align}
  \phi^\std_A &= \phi^\Std_A + V_N \ln \rho_A \label{eq:stdStd}\\
\text{where } \rho_A &= \frac{x_A^\std}{x_A^\Std} = \frac{c_A^\std}{c_A^\Std}
\end{align}
and $c_A^\std$ and $c_A^\Std$ are the concentrations at the relevant
conditions.

Whereas each \emph{species} \ch{A} is associated with a
\emph{potential} $\phi_A$, each \emph{reaction} $r$ is also associated
with a \emph{reaction potential} (which is denoted $\Phi$) split into
two components: the \emph{forward reaction potential} $\Phif$ and the
\emph{reverse reaction potential} $\Phir$.  The net reaction
potential, which drives the reaction, is given by
$\Phi = \Phif - \Phir$. In the case of the reaction (\ref{eq:ATP})
\begin{align}
  \Phif &= \phi_{\ch{ATP}} + \phi_{\ch{H2O}} \\
  \Phir &= \phi_{\ch{ADP}} + \phi_{\ch{HPO4}} + \phi_{\ch{H}}\\
  \Phi  &= \phi_{\ch{ATP}} + \phi_{\ch{H2O}}
 - \lb \phi_{\ch{ADP}} + \phi_{\ch{HPO4}} + \phi_{\ch{H}} \rb
\end{align}

The CR of a chemical \R component (assuming mass-action kinetics) is
\begin{align}
  f &= \kappa \lb \exp \frac{\Phif}{V_N} - \exp \frac{\Phir}{V_N} \rb\label{eq:f}
\end{align}
This CR requires the forward ($\Phif$) and reverse ($\Phir$) potentials separately; as
discussed by \citet{Kar90} this requires either an implicit modulation
or a two-port \R component.

\subsection{Numerical Values}
\label{sec:numerical-values}
Perhaps surprisingly, values for standard potentials and typical
cellular concentrations can be hard to find in the biochemical
literature. The values used in this paper come from two
sources. Chemical potentials $\mu^\Std$ at standard conditions are
taken from \citet[Table 5]{LiWuQi11} and converted to
Faraday-equivalent potentials $\phi^\Std=F\mu^\Std$. Concentrations
are taken \citet[Table 5]{ParRubXu16} and are used in conjunction with
Equation \eqref{eq:stdStd} to compute potentials at typical cellular
conditions.

Reconciling experimental data is a big issue that is beyond the scope
of this paper -- see, for example, \citet{TumKli18}. Using the
aforementioned data, some reactions discussed in
\S~\ref{sec:glycolysis} were found to have small negative potentials;
these are not thermodynamically feasible and so the data from
\citet[Table 5]{LiWuQi11} was modified to give small positive
potentials. In particular, the potentials of \ch{DHAP} and \ch{GAP}
were adjusted by \SI{10}{ mV}, about 0.1\%. In general, reaction
potentials are the difference of large species potentials and thus
small percentage changes in the latter can give large percentage
changes in the former.

Under such typical cellular conditions, reaction
(\ref{eq:ATP}) is associated with a Faraday-equivalent potential
$\Phi$ of about \SI{532}{ mV}; for this reason the reaction can be
used to pump chemical reactions against an adverse potential gradient.

\subsection{Redox reactions}
\label{sec:redox-reactions}
\begin{figure}[htbp]
  \centering
  \Fig{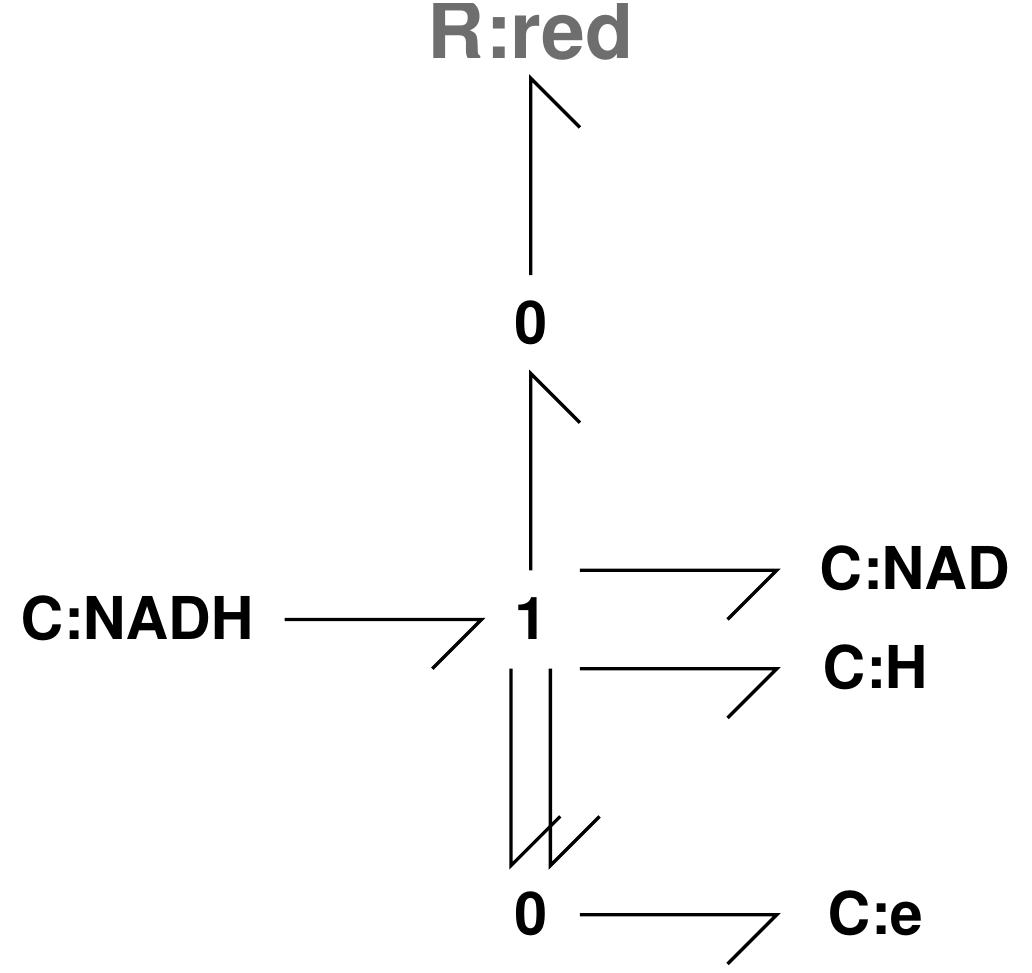}{\HalfWidth}
  \caption{\ch{NADH} reduction: reaction (\ref{eq:NADH})}
  \label{fig:NADH}
\end{figure}
Redox (reduction/oxidation) reactions are the key to aerobic life; the
bond graph modelling of such reactions is described by
\citet{Gaw17a}. Redox reactions can be split into two half reactions
each of which explicitly contains the electrons donated or accepted by
the reaction.  As an example of this in the first stage of the
mitochondrial electron transport chain, \ch{NADH} (reduced
Nicotinamide Adenine Dinucleotide) donates two \ch{e-} (electrons) in
forming \ch{NAD+} (oxidised Nicotinamide Adenine Dinucleotide) which
are accepted by {Q} (oxidised Ubiquinone) to form \ch{QH2} (reduced
Ubiquinone).
\begin{equation}\label{eq:NADH}
  \ch{ NADH <>[ red ]  NAD+ +  H+ + 2 e-}
\end{equation}
In Figure \ref{fig:NADH}, the chemical species and proton are modelled as in
\S~\ref{sec:chemical-equations}; the \ch{e-} is associated with the
\emph{redox potential} of the reaction and can be modelled by an
electrical capacitor \citep{Gaw17a}. Using the Faraday-equivalent
potential discussed in \S~\ref{sec:chemical-equations}, the potentials
of the chemical species are commensurate with the potentials of the
\ch{e-}. In particular, under typical cellular conditions
(\S~\ref{sec:numerical-values}), reaction (\ref{eq:NADH}) is associated
with a Faraday-equivalent potential $\Phi$ of about \SI{345}{ mV};
once again, for this reason the reaction can be used to pump chemical
reactions.

\subsection{Glycolysis}
\label{sec:glycolysis}
\begin{figure}[htbp]
  \centering
  \SubFig{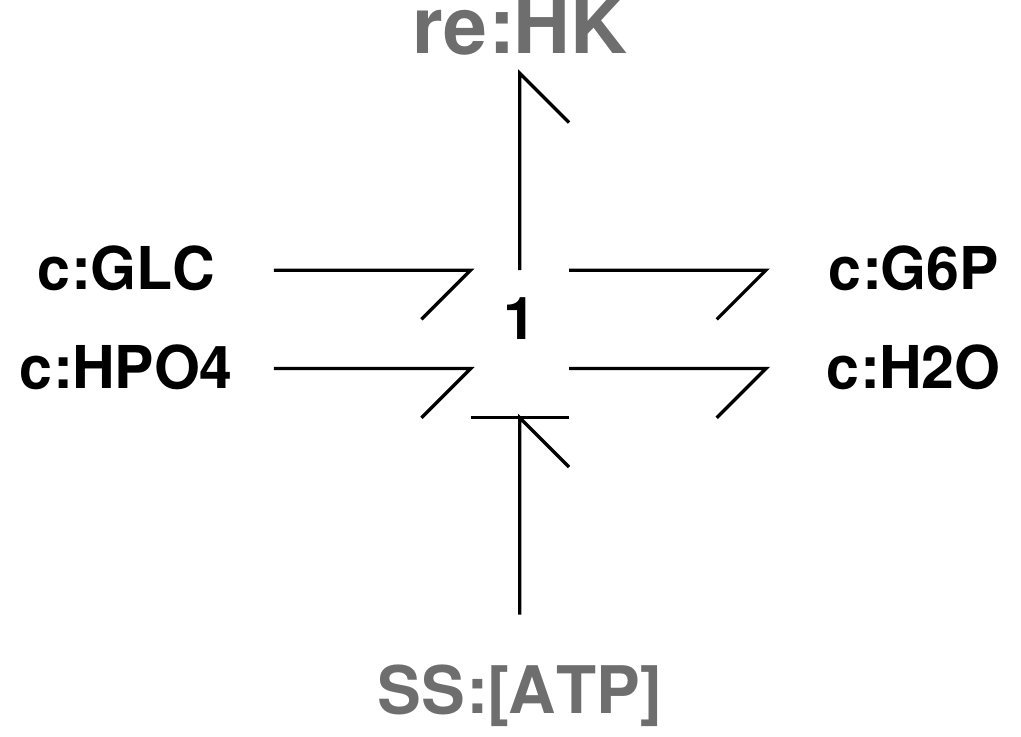}{Reaction}{\HalfWidth}
  \SubFig{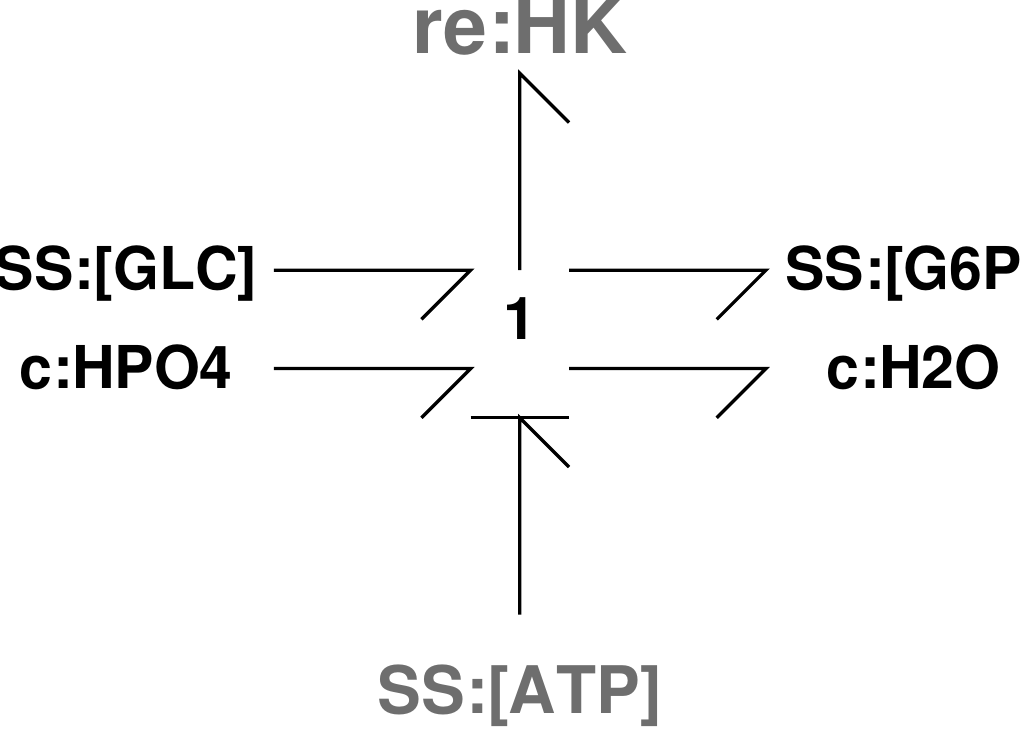}{Modular reaction}{\HalfWidth}
  \caption{Modularity: hexokinase (\ch{HK}) reaction}
  \label{fig:HK}
\end{figure}
The enzyme hexokinase is involved in a reaction which converts glucose
(\ch{GLC}) to glucose 6-phosphate (G6P):
\begin{equation}
  \label{eq:G6P}
  \ch{ ATP +  GLC <>  ADP +  H +  G6P}
\end{equation}
This can be rewritten as the combination of two reactions: \ch{ATP}
hydrolysis reaction (\ref{eq:ATP}) and
\begin{equation}
  \label{eq:HK}
  \ch{ GLC + HPO4 <>[HK]  G6P + H2O}
\end{equation}
As in \S~\ref{sec:chemical-equations}, the HK reaction (\ref{eq:HK})
has the bond graph representation of Figure \ref{subfig:hk_abg} where
the bond graph \emph{source-sensor} component \BSS{[ATP]} provides a
port to connect to the \ch{ATP} hydrolysis reaction (\ref{eq:ATP}).
As  the HK reaction (\ref{eq:HK}) is to be embedded in a larger model,
a modular version is obtained by replacing \BC{GLC} and \BC{G6P} by the
ports \BSS{[GLC]} and \BSS{[G6P]} respectively.

\begin{figure}[htbp]
  \centering
  \SubFig{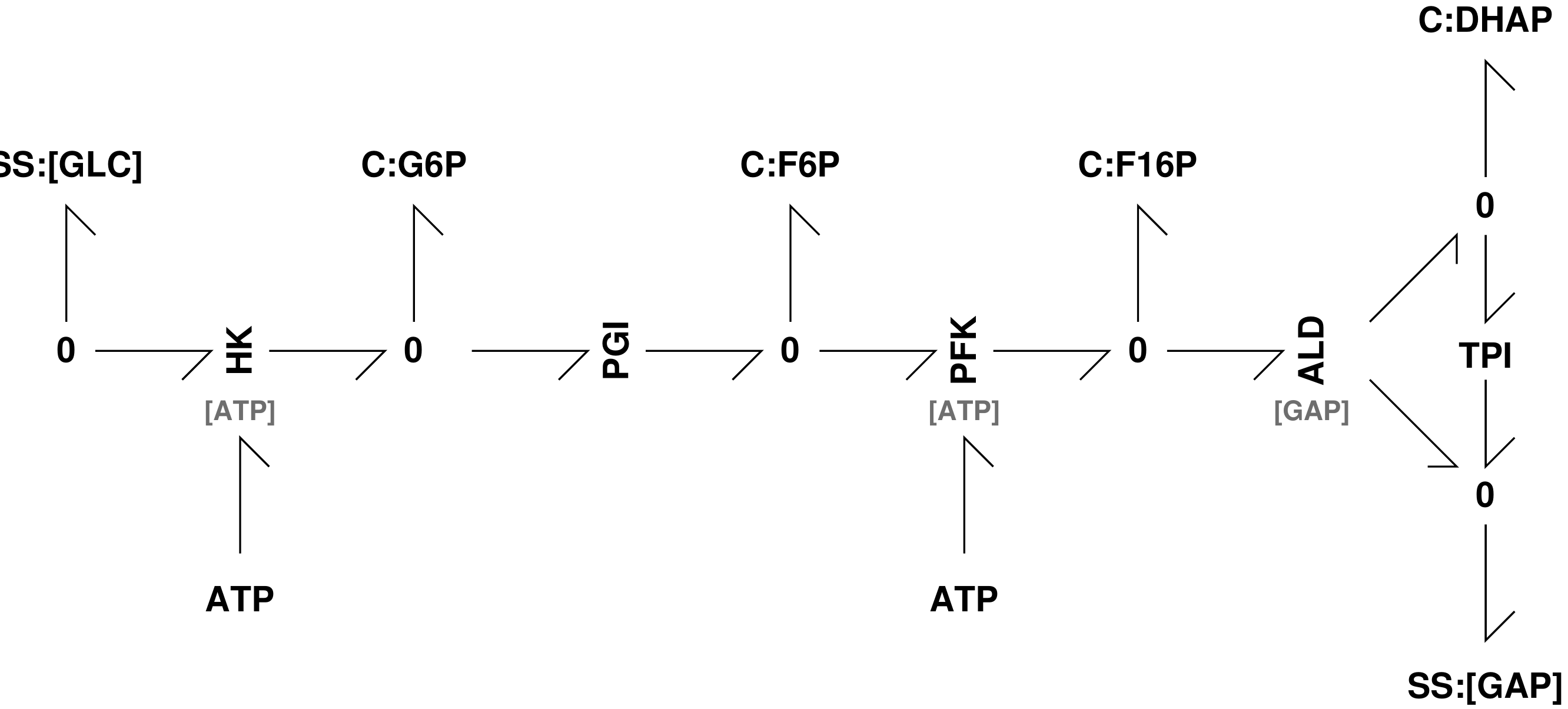}{GlyA: Stage 1}{\FullWidth}
  \SubFig{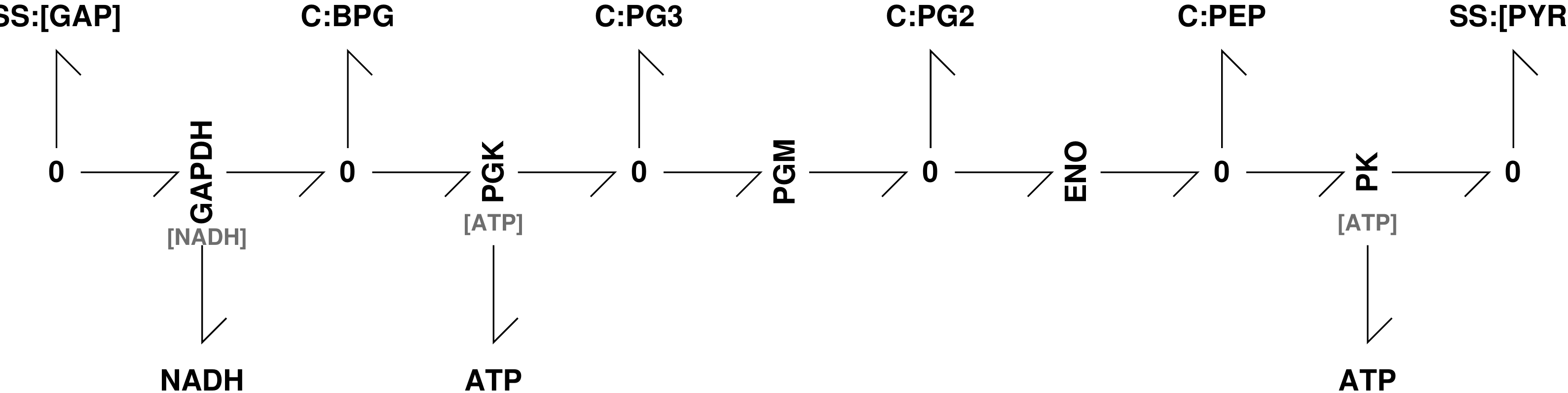}{GlyB: Stage 2}{\FullWidth}
  \SubFig{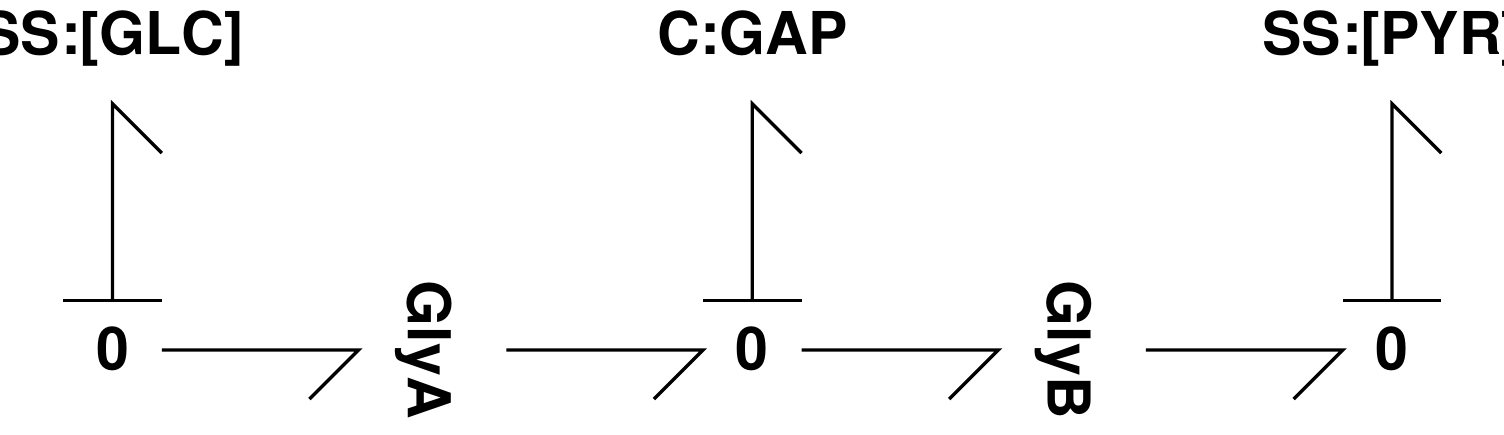}{Gly}{\FullWidth}
  \caption{Glycolysis.}
  \label{fig:Gly}
\end{figure}
The bond graph  models of the two stages of glycolysis are given in
Figures \ref{subfig:GlyA_abg} and \ref{subfig:GlyB_abg} and are
combined in Figure \ref{subfig:Gly_abg}.

 Figure \ref{subfig:GlyA_abg} shows the modular version of HK embedded
 in the first stage of glycolysis. This clearly shows the two key
 features of the HK catalysed reaction: it converts converts GLC to
 G6P and it is pumped by ATP hydrolysis. 
 In Figure \ref{subfig:GlyA_abg}, the two pathways diverging from the
 \ch{ALD} reaction converge on \ch{GAP} indicate that stage 1 of
 glycolysis converts each molecule of \ch{GLC} to two molecules of
 glyceraldehyde 3-phosphate (\ch{GAP}) and is pumped by two ATP
 hydrolysis reactions \eqref{eq:ATP}.
 Figure \ref{subfig:GlyB_abg} shows that stage 2 of glycolysis
 converts each molecule of \ch{GAP} to one molecule of \ch{PYR} and,
 as indicated by the bond arrow direction, pumps two reverse ATP
 hydrolysis reactions \eqref{eq:ATP} and a reverse \ch{NADH} reaction
 \eqref{eq:NADH}.

 The modular bond graph of Figure \ref{fig:Gly} is equivalent to the
 biomolecular system where the associated reaction potentials $\Phi$
 correspond to typical cellular conditions
 (\S~\ref{sec:numerical-values}).

 {\scriptsize
\begin{xalignat}{2}
\Ch{ GLC +  ATP &<>[ HK ]  G6P +  ADP +  H} & &(\SI{241}{ mV})\notag\\
\Ch{ G6P &<>[ PGI ]  F6P} & &(\SI{70}{ mV})\notag\\
\Ch{ F6P +  ATP &<>[ PFK ]  F16P +  ADP +  H} & &(\SI{135}{ mV})\notag\\
\Ch{ F16P &<>[ ALD ]  GAP +  DHAP} & &(\SI{32}{ mV})\notag\\
\Ch{ DHAP &<>[ TPI ]  GAP} & &(\SI{10}{ mV})\notag\\
\Ch{ GAP +  HPO4 +  NAD + 2 e &<>[ GAPDH ]} &&\notag\\ 
&\Ch{H +  BPG +  NADH} &&(\SI{14}{ mV})\notag\\
\Ch{ ADP +  BPG &<>[ PGK ]  ATP +  PG3} & &(\SI{39}{ mV})\notag\\
\Ch{ PG3 &<>[ PGM ]  PG2} & &(\SI{30}{ mV})\notag\\
\Ch{ PG2 &<>[ ENO ]  H2O +  PEP} & &(\SI{27}{ mV})\notag\\
\Ch{ ADP +  H +  PEP &<>[ PK ]  ATP +  PYR} & &(\SI{65}{ mV})\notag
\end{xalignat}
}
%





\section{EFFICIENCY}
\label{sec:efficiency}
\begin{figure}[htbp]
  \centering
  \Fig{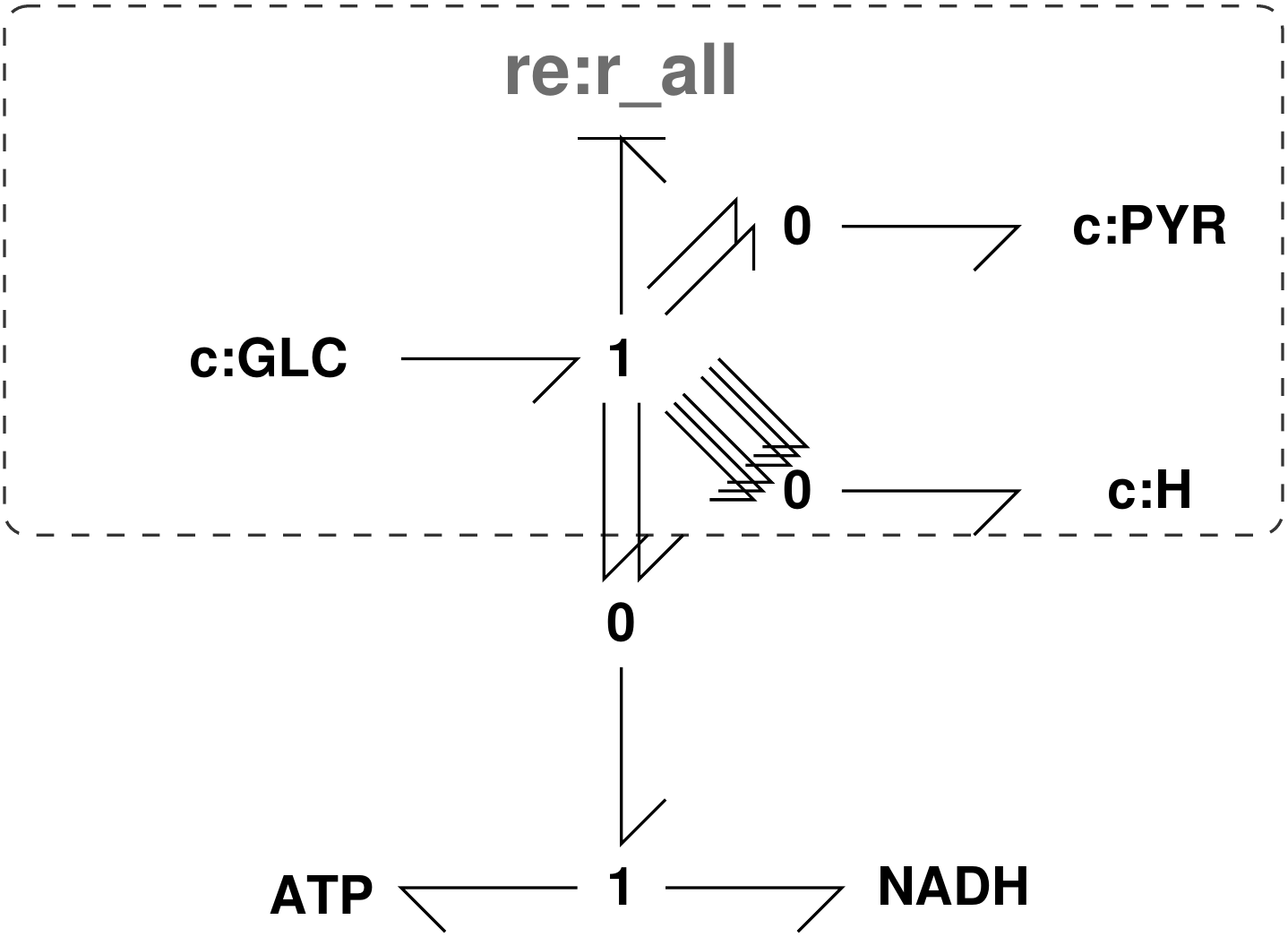}{\FullWidth}
  \caption{Pumping. The reaction \ch{GLC <> 2 PYR + 6 H} pumps the
    reverse \ch{ATP} hydrolysis reaction of Figure \ref{fig:ATP} and
    the reverse \ch{NADH} reduction reaction of Figure \ref{fig:NADH}
    represented by the modules \textbf{ATP} and \textbf{NADH}
    repectively.}
  \label{fig:GlyPump}
\end{figure}
The biomolecular network implementing glycolysis discussed in
\S~\ref{sec:glycolysis} converts  glucose (\ch{GLC}) to pyruvate
(\ch{PYR}) as well as driving \ch{ATP} hydrolysis and the reduction of
\ch{NADH} in reverse to store chemical energy. In particular, the
overall reaction is:
{\scriptsize
\begin{xalignat}{2}
\ch{ GLC + 2 ADP + 2 HPO4 + 2 NAD + 4 e &<>[ r_{all} ]&&\\
  2 ATP + 2 H + 2 H2O + 2 NADH + 2 PYR}& & &(\SI{837}{ mV})\notag
\end{xalignat}
} At the standard conditions discussed in
\S~\ref{sec:numerical-values} and denoted by the $^\std$ symbol, this
reaction is associated with a reaction potential
\begin{equation}
  \Phi^\std_{all} = \SI{837}{ mV}
\end{equation}
and the corresponding power dissipation is:
\begin{equation}
  P^\std_{diss} = \Phi^\std_{all}f \si{pW}
\end{equation}
where $f\si{mA}$ is the reaction flow rate per unit volume.

As in Figure \ref{fig:GlyPump}, this reaction can be split into three parts:
{\scriptsize
\begin{xalignat}{2}
\ch{ GLC &<>  2 PYR + 6 H} & &(\SI{2712}{ mV})\label{eq:GLC-PYR}\\
\ch{ 2 ADP +  2 H +  2 HPO4 &<> 2 ATP +  2 H2O  } & &(\SI{-1184}{
  mV})\notag\\
\ch{2  NAD +  2 H + 4 e &<> 2 NADH } & &(\SI{-690}{ mV})
\end{xalignat}
} 

The latter two reactions represent two reverse \ch{ATP} hydrolysis
reaction \eqref{eq:ATP} and two reverse \ch{NADH} reduction reactions
\eqref{eq:NADH} respectively; the first reaction represents the
remainder of the reaction converting \ch{GLC} to \ch{PYR} (and
\ch{H+}).

The first reaction is associated with a reaction \emph{driving potential}
\begin{equation}\label{eq:Phi_0}
  \Phi^\std_0 = \phi^\std_{GLC} - 2\phi^\std_{PYR} - 6\phi^\std_{H} = \SI{2712}{ mV}
\end{equation}
Because the latter two reactions are being pumped by the first, define
the \emph{pumping potential} $\Phi_{pump}$ as:
\begin{align}
  \Phi^\std_{pump} &= 2 \lb \Phi^\std_{ATP} + 2 \Phi^\std_{NADH}\rb\notag\\
              &= 1184 + 690 = \SI{1874}{ mV}
\end{align}
These potentials are associated with a corresponding \emph{driving
  power} $P_{0}$ and \emph{pumping power} $P_{pump}$:
\begin{align}
  P_{0} &= \Phi_{0}f\\
  P_{pump} &= \Phi_{pump}f
\end{align}

With this example in mind define the \emph{pumping efficiency} as the
ratio of pumping power to driving power
\begin{align}
  \eta &= \frac{P_{pump}}{P_0}\notag\\
  &= \frac{\Phi_{pump}}{\Phi_0}\label{eq:eta}
\end{align}
At standard conditions
\begin{align}
\eta^\std= \frac{\Phi^\std_{pump}}{\Phi^\std_0} = \frac{1874}{2712} = 69.1\% \label{eq:eta_std}
\end{align}

\begin{figure}[htbp]
  \centering
  \SubFig{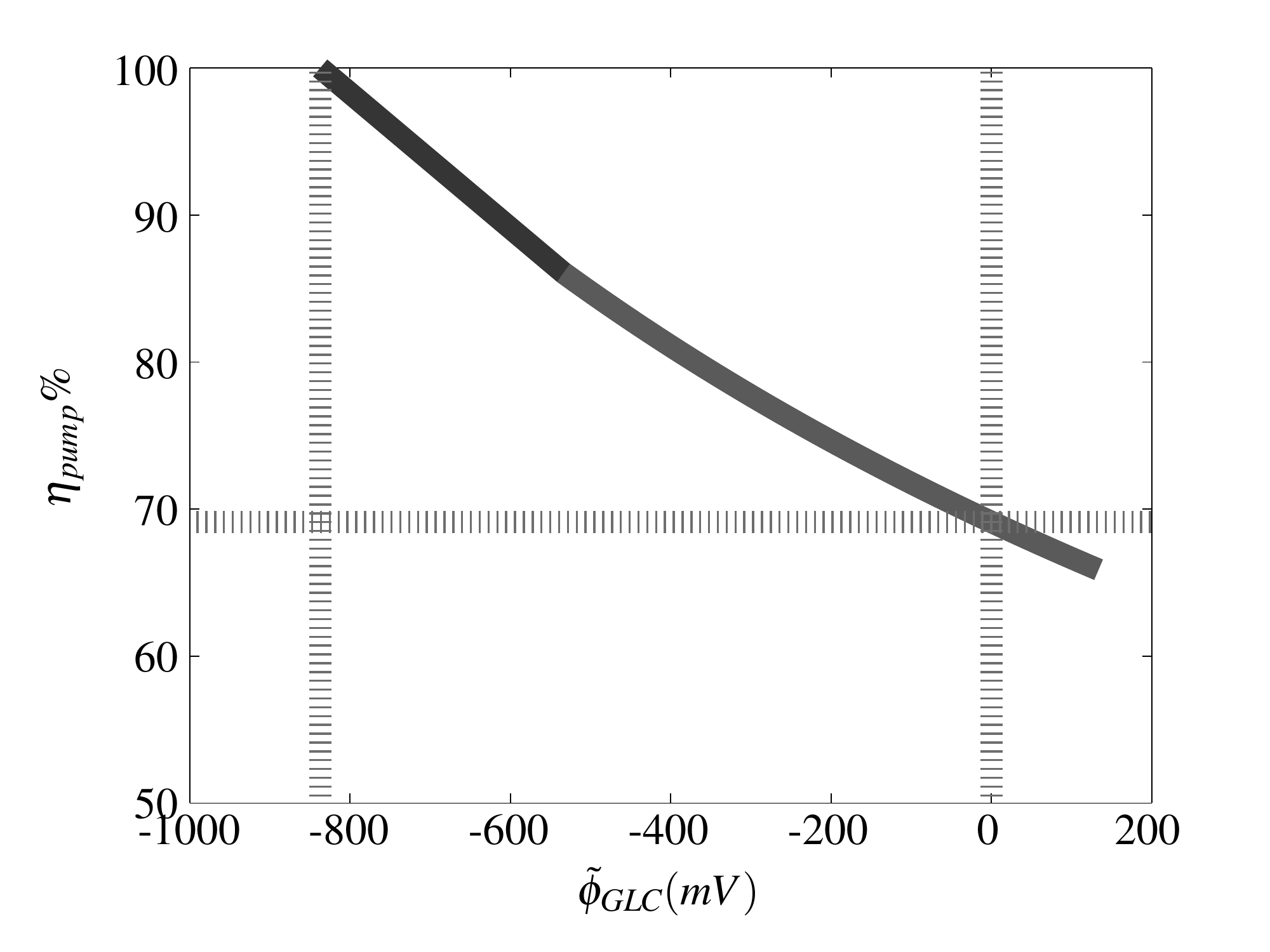}{}{\FullWidth}
  \SubFig{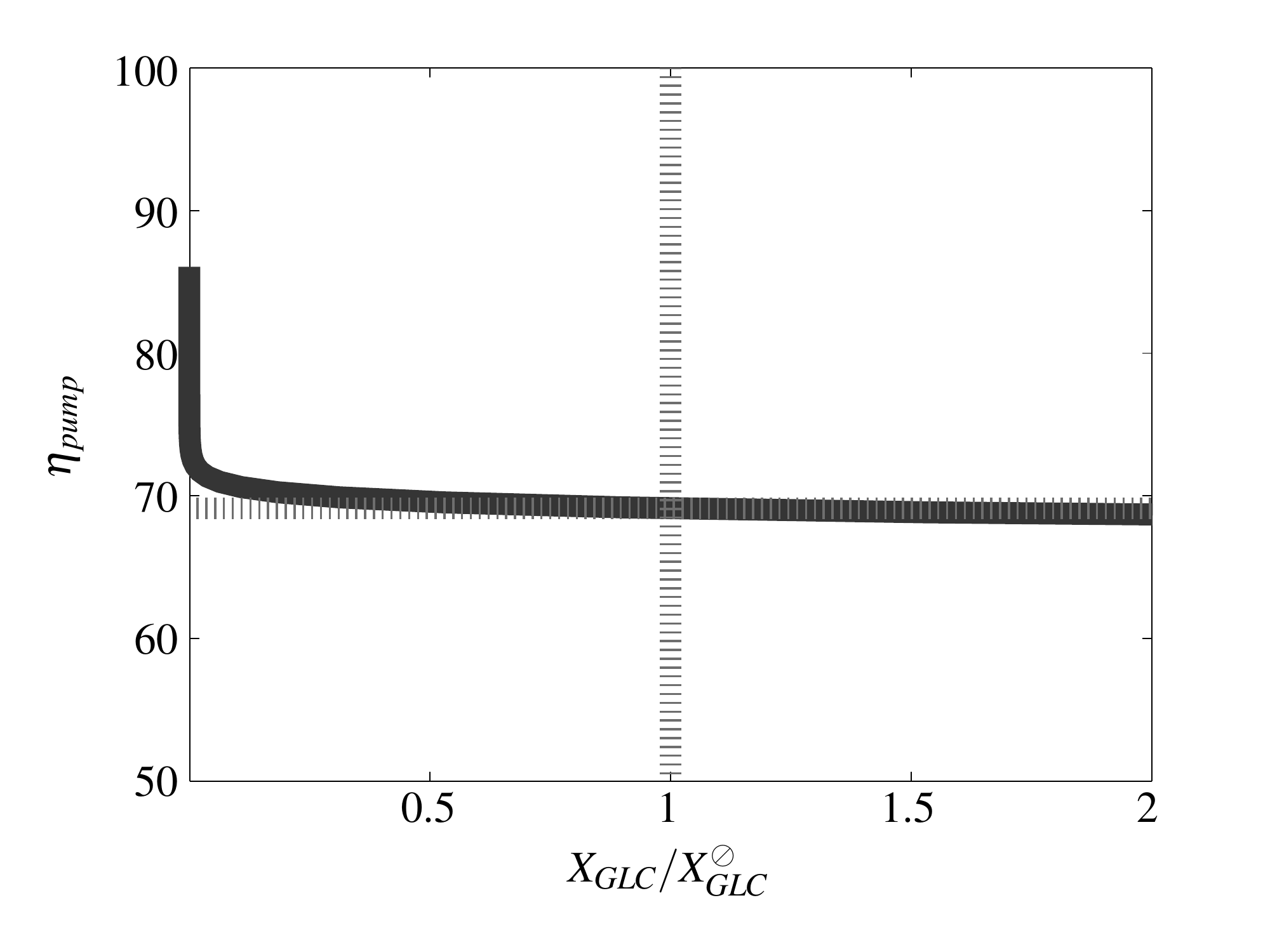}{}{\FullWidth}
  \caption{Pumping Efficiency $\eta_{pump}$. Nominal conditions are
    indicated by the broken lines.
    (a) Plotted against normalised potential of \ch{GLC}:
    $\tilde{\phi}_{GLC}$.
    (b) Plotted against normalised concentration of \ch{GLC}.
   }
  \label{fig:eta}
\end{figure}

The efficiency computed in Equation \eqref{eq:eta_std} corresponds to
the nominal values discussed in \S~\ref{sec:numerical-values}, which
in turn correspond to the nominal flow of $f^\std=\SI{2.3}{mM/min}$.
If the concentration of glucose (\ch{GLC}) is varied, the corresponding
potential $\phi_{GLC}$ varies according to Equation \eqref{eq:phi_A}.
then so will $\Phi_0$ of Equation \eqref{eq:Phi_0} and the pumping
efficiency \eqref{eq:eta}.
\begin{align}
  \eta &= \frac{\Phi^\std_{pump}}{\Phi^\std_0 + \tilde{\phi}_{GLC}}\notag\\
  \where \tilde{\phi}_{GLC} &= \Phi_0 - \Phi^\std_0 = \phi_{GLC} - \phi^\std_{GLC} 
\end{align}
Figure \ref{subfig:GLC_eta}  shows how $\eta$ varies with
$\tilde{\phi}_{GLC}$. Note that
$\tilde{\phi}_{GLC}=\Phi^\std_{pump}-\Phi^\std_0 = -\Phi^\std_{all}$
corresponds to $\eta=100\%$. As this value of $\tilde{\phi}_{GLC}$
corresponds to $\Phi^\std_{all}=0$, the flow $f=0$ at this point.

Figure \ref{subfig:GLC_X_eta} shows efficiency $\eta$ plotted against
the normalised concentration of \ch{GLC}.

\begin{figure}[htbp]
  \centering
  \Fig{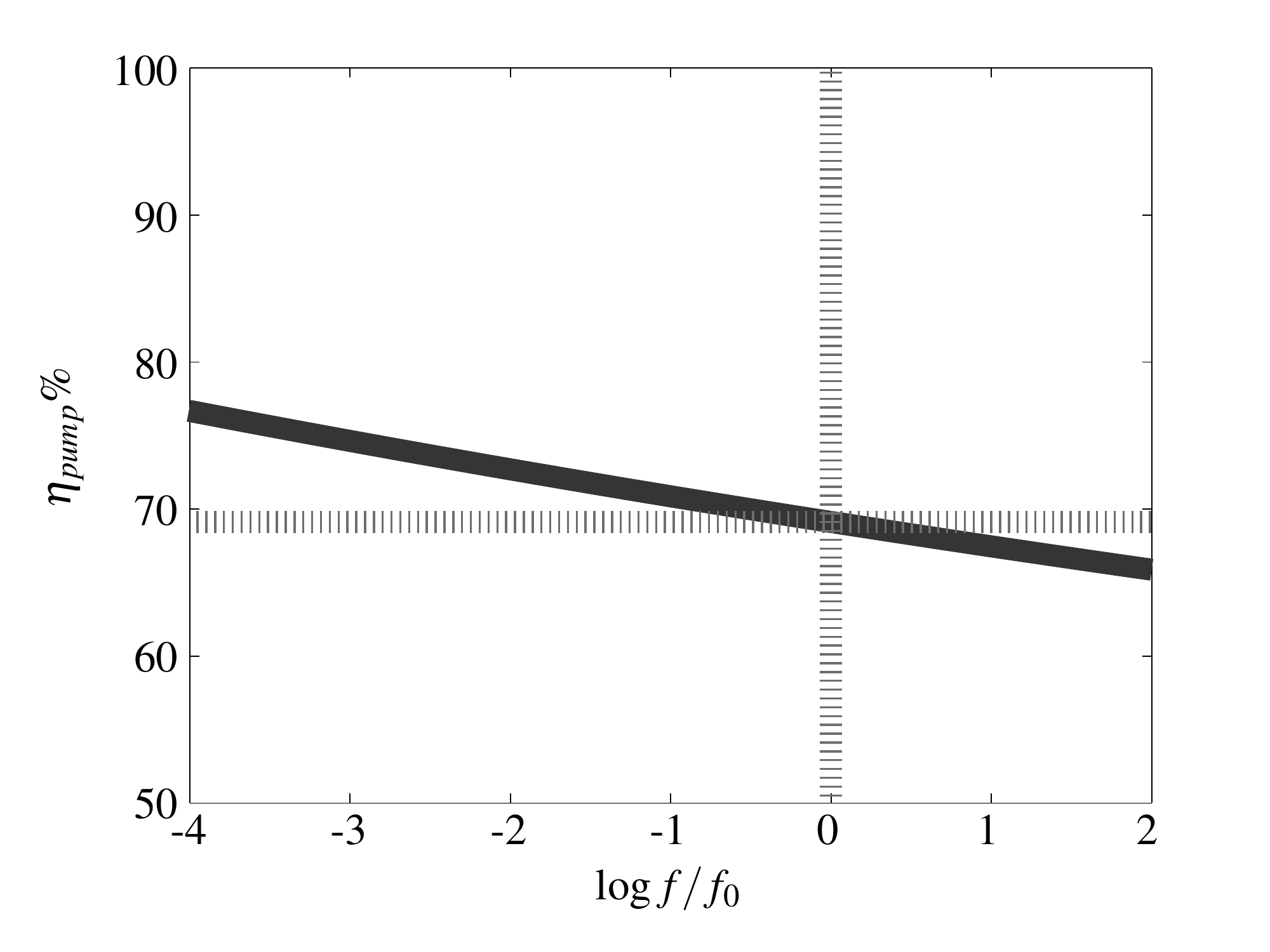}{\FullWidth}
  \caption{Efficiency $\eta$ plotted against log normalised flow:
    $\log_{10} f/f_0$.  }\label{fig:eta_flow}
\end{figure}

The computation generating the data for Figure \ref{fig:eta} does not
involve the Faraday-equivalent flow $f$. However, efficiency as a
function of flow is of interest. As a approximation to this, the
steady-state flow $f$ was computed as $\phi_{\ch{GLC}}$ was varied
assuming that all reactions have the mass-action kinetics of
\eqref{eq:f} using the method of \citet{Gaw18}. This was used to
generate the data for Figure \ref{fig:eta_flow}. In fact, the reaction
kinetics are more complicated than the mass-action representation
hence the computations are more challenging than those used to
generate Figure \ref{fig:eta_flow}.

These results indicate that, under these particular conditions, the
pumping efficiency of glycolysis is around 70\% except for very low
flow rates associated with low concentrations of Glucose (\ch{GLC})
and thus $\Phi_0$ being only slightly larger that $\Phi_{pump}$.

%

\section{CONCLUSION}
\label{sec:conclusion}
The basic ideas of modelling biomolecular systems using bond graphs
and the Faraday-equivalent potential have been outlined and
illustrated using the example of glycolysis: the first stage of
aerobic respiration.

The concept of \emph{pumping efficiency} has been introduced and
illustrated using glycolysis and experimental numerical values drawn
from the recent paper of \citet{ParRubXu16}. These ideas are currently
being extended to mitochondrial metabolism: the TCA cycle and the
electron transport chain.


\end{document}

%% file: BG-sym.tex
\newcommand{\BG}[1]{\text{\sffamily\textbf{#1}}}

\newcommand{\C}{\BG{C }}

\newcommand{\R}{\BG{R }}

\newcommand{\BGL}[2]{$\BG{#1}$:$\mathbf{#2}$} 

\newcommand{\BC}[1]{\BGL{C}{#1}}

\newcommand{\BSS}[1]{\BGL{SS}{#1}}


%% file: BG-maths.tex
\usepackage{amsmath,amssymb,amscd,amstext,mathtools,extarrows,centernot}

\newcommand{\lb}{\left (}
\newcommand{\rb}{\right )}

\newcommand{\where}{\text{where }}

\newcommand{\Phif}{{\Phi^f}}
\newcommand{\Phir}{{\Phi^r}}

\usepackage{listings}


%% file: BG-chem.tex
\usepackage{chemformula}

\newcommand{\Std}{\ominus}
\newcommand{\std}{\oslash}

